\documentstyle[12pt,epsfig]{article}
\begin{document}
\begin{flushright}
KIAS--P01020\\
LPMT--01-21\\
April 2001 \\
\end{flushright}
\newcommand{\nn}{\noindent}
\renewcommand{\thefootnote}{\fnsymbol{footnote}}
\begin{center}
{\Large{ {\bf Supersymmetric enhancement of associated $ZA^0$ production
at $e^+e^-$ colliders
}}}

\vspace{1.6cm}

{\large A.G. Akeroyd}$^{\mbox{1}}$,
{\large A. Arhrib}$^{\mbox{2}}$\footnote{On leave of 
absence from Department of Mathematics, FSTT, P.O.B416 Tangier, Morocco.},
{\large M. Capdequi Peyran\`ere}$^{\mbox{3}}$

\vspace{1.3cm}
{\sl
1: Korea Institute for Advanced Study, 207-43 Cheongryangri--dong, \\
Dongdaemun--gu, Seoul 130--012, Korea

\vspace{.6cm}
2: Physics Department, National Taiwan University 10764 Taipei, Taiwan, R.O.C 

\vspace{.6cm}
3: Laboratoire de Physique Math\'ematique et Th\'eorique, 
CNRS-UMR 5825\\
Universit\'e Montpellier II, F--34095 Montpellier Cedex 5, France }
\end{center}

\setcounter{footnote}{4}
\vspace{1.8cm}

\begin{abstract}
\nn We study the associated production of the 
$A^0$ neutral CP--odd Higgs boson 
with  a neutral gauge boson $Z$
in high energy $e^+ e^-$ collisions at the one loop level.  
We present a detailed discussion for the total cross--section 
predicted in the context of the Minimal Supersymmetric 
Standard Model (MSSM) and make a comparison with the non--SUSY 
Two Higgs Doublet Model (THDM). We show that the MSSM cross-section 
may be enhanced compared to that for the THDM. 
\end{abstract}
\vfill

\newpage

\pagestyle{plain}
\renewcommand{\thefootnote}{\arabic{footnote} }
\setcounter{footnote}{0}

\section*{1.~Introduction}

The Standard Model (SM) of electroweak
interactions \cite{Wein} is in complete agreement
with all precision experimental data (LEP, Tevatron, SLD).
So far undiscovered is the SM Higgs boson ($\phi^0$), 
which is  the major goal of present and future 
colliders. Global SM fits \cite{osaka2} 
favour a relatively light $\phi^0$, 
suggesting that it should be observed soon at forthcoming colliders. 
A light Higgs boson is also a feature 
of Supersymmetric (SUSY) theories,
which are the most popular extensions of the SM.
Since 1989 LEP has searched for $\phi^0$ in the energy range 
91 GeV $\le \sqrt s\le 208$
GeV via the mechanism $e^+e^-\to Z^{(*)}\phi^0$. 
In the final run at energies around $\sqrt s=208$ GeV  
two of the four experiments
found excesses in the search for Higgs boson production \cite{evid}
in association with a Z boson (Higgsstrahlung), while the other two presented
improved lower limits on its mass \cite{noevid} of $M_{\phi^0}> 109.7$
GeV and $>114.3$ GeV.

LEP agreed that a further run with about 200 pb$^{-1}$ per experiment at
$\sqrt s=208.2$ GeV would be enable the combined data 
from the four experiments to
establish a 5$\sigma$ discovery \cite{www}. However the extended 
run was not approved and LEP was consequently shut down. 
The Higgs search will continue with Run II at the Tevatron 
\cite{tevatronH} which has a chance of confirming the existence 
of the Higgs boson in the mass range hinted at by 
LEP ($M_{\phi^0}\approx 115$ GeV). This region is fairly 
problematic for the LHC \cite{LHC}
and would require several years 
searching for $\phi^0\to \gamma\gamma$ decays 
in order to confirm such a light Higgs.

The still hypothetical Higgs sector of the Standard Model 
(SM) can be  enlarged and some simple extensions 
such as the Minimal Supersymmetric Standard Model 
(MSSM) and Two Higgs Doublet Model (THDM) \cite{HHG, MSSM} 
are under much intensive study. 
Both the THDM and MSSM introduce 2 Higgs doublets which 
break the electroweak symmetry \cite{ewsb}. From 
the 8 degrees of freedom initially present in the 
2 Higgs doublets, 3 correspond to the masses of the 
longitudinal gauge bosons, 
leaving 5 degrees of freedom which manifest themselves as 5 
physical Higgs particles (2 charged Higgs $H^\pm$, 
2 CP--even $H^0$, $h^0$ and one CP--odd $A^0$). 
Aside from the charged Higgs sector, 
the neutral scalars of the THDM may possess a very different
phenomenology to that of the SM Higgs. 
The discovery  of a CP--odd Higgs boson or/and 
charged Higgs boson would be clear evidence of 
physics beyond the SM. Detection of a CP--even neutral 
Higgs with couplings differing significantly from
those expected for $\phi^0$ \cite{beySM} would 
also provide evidence for physics beyond the SM.
If the Higgsstrahlung cross--section is suppressed then
the mass bound on such a Higgs boson is much weaker than the 
above bounds on $M_{\phi^0}$.

Until now, no Higgs boson has been discovered, 
and from negative searches one can
derive direct and indirect bounds on their masses. 
The combined null--searches from all four CERN 
LEP collaborations derive the 
lower limits $M_{H^{\pm}}\ge 77.3$ GeV and $M_A\ge 90$ GeV
\cite{www,osaka} in the context of the MSSM.

CP-odd Higgs bosons can be produced at $e^+e^-$ colliders 
\cite{eecolliders} via $e^+e^- \to h^0 A^0$ and 
$e^+e^- \to b\overline bA^0,t\overline tA^0$ \cite{eeA},\cite{Dawson}.
At future $e^+e^-$ colliders the simplest way to produce 
CP--even Higgs scalars is in the Higgsstrahlung process 
$e^+e^-\to Z^*\to HZ$. The CP--odd $A^0$ possesses no
tree-level coupling $A^0$ZZ, and the other tree--level diagrams 
for such a process are proportional
to the electron mass and consequently negligible
\footnote{Note that at a muon collider \cite{aad}, 
the tree--level diagrams 
cannot be discarded anymore and higher order diagrams 
would induce corrections to the tree-level rate.}. The
dominant contribution is therefore from higher order diagrams
which will be mediated by both SM and non--SM particles.
Therefore the rates are expected to be strongly model dependent.

Previous work on the loop induced production of Higgs bosons
in association with gauge bosons (e.g. $e^+e^-\to A^0\gamma,H^{\pm}W^{\mp}$)
can be found in \cite{abdel,achm,AAC} and references therein.
Recently, associated production $\gamma\gamma \to ZH$, 
both for SM Higgs and MSSM Higgs, has been studied in \cite{fernand}. 
In a previous paper \cite{AAC} we studied the 
process $e^+e^-\to A^0Z$, in the context of the THDM. 
In this paper we extend the study to the case of the 
MSSM. The particle spectrum of the 
MSSM doubles that of the THDM, but the scalar sectors
are equivalent up to some mass relations and differences
in the couplings \cite{HHG}. One would expect differences
in the rates for the THDM and the MSSM, since in the latter 
there will be contributions from loops involving 
sparticles. If $e^+e^-\to Z^*\to A^0Z$ 
were sizeable it would provide an alternative 
way of producing $A^0$ at $e^+e^-$ colliders, 
with a kinematical reach superior to that
for the mechanism $e^+e^-\to Z^*\to A^0H^0$. 
Note that in the MSSM the kinematically favoured mechanism 
$e^+e^-\to Z^*\to A^0h^0$ is suppressed by the factor
$\cos^2(\beta-\alpha)$ which is very small in the 
region $M_A\ge 200$ GeV. The study of the various 
production mechanisms of the CP--odd Higgs boson is 
well motivated since the discovery of such a 
particle would signify that the electroweak 
symmetry breaking is introduced by more that one 
Higgs doublet ($\phi^0$ is CP--even).

Our work is organized as follows. In section 2 we outline our
approach for evaluating the 1--loop rate for $e^+e^-\to A^0Z$
in the MSSM. In section 3 we present our numerical results and section 4 
contains our conclusions.

\begin{figure}[t!]
\smallskip\smallskip 
\centerline{{
\epsfxsize4.8 in 
\epsffile{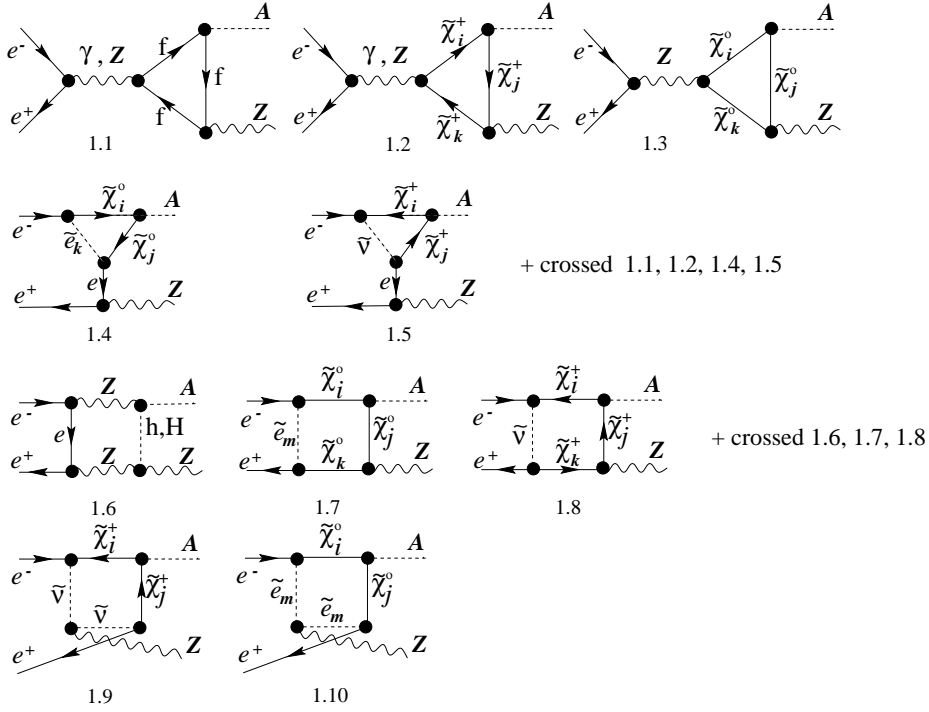}}}
\smallskip\smallskip
\caption{ Generic Feynman diagrams (vertices and boxes) 
 contributing to $e^+e^-\to Z A^0$ in the MSSM}
\label{fey}
\end{figure}

\renewcommand{\theequation}{2.\arabic{equation}}
\setcounter{equation}{0}
\section*{2. One--Loop Corrections}
In the limit of vanishing electron mass, the process 
$e^+e^- \to Z A^0$ posseses no tree--level contribution 
and is thus mediated by higher order diagrams. 
We have evaluated the one--loop amplitude in the 
'tHooft--Feynman gauge.
This amplitude contains ultraviolet divergences (UV) 
and we will use the dimensional regularization scheme \cite{thooft} 
to deal with them.\\
The typical Feynman diagrams for the virtual corrections 
of order $\alpha^2$ are drawn in Fig.1. These comprise: 
\begin{itemize}
\item [{(i)}] The fermionic 
contribution (s--channel) to
the vertices $\gamma$-Z-$A^0$ and Z-Z-$A^0$: 
Fig.1.1 $\to$ 1.2 (+ crossed counterpart) and Fig.1.3 
\item [{(ii)}] SUSY t--channel correction to the vertex $e^-e^{-*}A^0$: 
Fig.1.4, Fig.1.5 (+ crossed counterparts). 
This kind of vertex is non--vanishing in the limit of 
vanishing electron mass because one of the electrons is 
off--shell. 
\item [{(iii)}] THDM box contributions, Fig. 1.6, and MSSM box
contributions, Fig. 1.7 $\to$ 1.8, (+ crossed
counterparts) and topology Fig. 1.9 and Fig. 1.10.
Note that one--loop contributions
coming from initial state $e^+ e^- H_0$,
$e^+ e^- h_0$ vertices also vanish in the limit $m_e\to 0$, since
$e^+$ and $e^-$ are both on--shell.
\end{itemize}

The t--channel vertex (Fig.1.5) and box diagrams 
Fig.1.8 and Fig.1.9 with 
exchange of charginos/sneutrinos contain vertices 
which violate fermion--number and
we use Denner rules \cite{drule} to deal with them.
All the Feynman diagrams are generated and
computed using FeynArts and FeynCalc \cite{seep} 
packages. We also use  the fortran FF--package 
\cite{ff} in the numerical analysis. 

In the MSSM, the s--channel topology in (i) also receives contributions 
from sfermion triangular loops. However, the coupling of $A^0$ 
to a pair of sfermions satisfies the following relation:
$(A^0\tilde{f}_i \tilde{f}^*_j)_{i\neq j} = 
-(A^0\tilde{f}_j \tilde{f}^*_i)_{i\neq j}$
and consequently the set of sfermions contributions to 
$\gamma$-Z-$A^0$ and Z-Z-$A^0$ vanishes. \\
Diagrams which involve Z-$A^0$ 
(resp Z-$G^0$) mixing in the s--channel self--energy and external lines
vanish because the vertices $\gamma$-$A^0$-$A^0$ (resp $\gamma$-$G^0$-$A^0$)
and Z-$A^0$-$A^0$ (resp Z-$G^0$-$A^0$) are absent. 
Z-$A^0$ mixing has to be considered in 
the t--channel, but owing to Lorentz invariance 
the Z-$A^0$ self--energy is proportional to CP odd 
momentum $p_{A^0}^{\mu}=(p_{e^+}+p_{e^-}-p_Z)^{\mu}$; 
then, since the vector boson Z is on--shell,
the t--channel amplitude will be
proportional to $m_e$ and consequently vanishes. 
Note also that the 
tadpole diagrams have vanishing contributions. 
We stress in passing that, there are diagrams in the process 
$e^+e^-\to A^0Z$ which do not contribute to 
$e^+e^-\to A^0\gamma$. These are diagrams
which involve the couplings $Z\chi^0\chi^0$ and 
$Z\tilde \nu \tilde \nu$, which would be absent if 
$Z$ is replaced by $\gamma$.

At the one--loop order and in the limit of vanishing 
electron mass, the amplitude ${\cal M}^1$ fully projects 
onto six invariants, in a similar way as in the 
THDM \cite{AAC}, as follows:
$$ {\cal M}^1 =\sum_{i=1}^{6} {\cal M}_i {\cal A}_i$$
where the invariants ${\cal A}_i$ are given by:
\begin{eqnarray}
& & {\cal A}_1 = \bar{v}(p_{e^+}) \not\!\epsilon (p_Z) \frac{1+\gamma_5}{2}
u(p_{e^-})\qquad\quad\ \ \ \ , \quad 
{\cal A}_2 = \bar{v}(p_{e^+}) \not\!\epsilon (p_Z) \frac{1-\gamma_5}{2}
u(p_{e^-})\label{ai}\\ & &
{\cal A}_3 =\bar{v}(p_{e^+}) \not\! p_Z \frac{1+\gamma_5}{2} u(p_{e^-})
(p_{e^-}\epsilon(p_Z))\nonumber \quad , \quad
{\cal A}_4=\bar{v}(p_{e^+}) \not\! p_Z
\frac{1-\gamma_5}{2} u(p_{e^-}) (p_{e^-}\epsilon
(p_Z))\nonumber \\ & &
{\cal A}_5=\bar{v}(p_{e^+}) \not\! p_Z
\frac{1+\gamma_5}{2} u(p_{e^-}) (p_{e^+}\epsilon
(p_Z))\nonumber \quad , \quad
{\cal A}_6=\bar{v}(p_{e^+}) \not\! p_Z
\frac{1-\gamma_5}{2} u(p_{e^-}) (p_{e^+}\epsilon
(p_Z)) \label{inva}
\end{eqnarray}
Here $\epsilon$ is the polarization vector of
the gauge boson Z.  
Summing over the Z gauge boson polarizations,
the squared amplitude may be written as:
\begin{eqnarray}
 \sum_{Z Pol} \vert {\cal M}^1 \vert^2 & = &
2 s (\vert {{\cal M}_1 \vert }^2 + \vert{{\cal M}_2 \vert }^2) -
\frac{(M_{A}^2M_Z^2 - t u)}{4 M_Z^2}
    \{ 4  \vert {{\cal M}_1 \vert }^2 + 4 \vert {{\cal M}_2 \vert }^2
 +\nonumber \\ & &  4 (M_Z^2 - t) Re[
{{\cal M}_1}{{\cal M}^*_3} +  {{\cal M}_2}{{\cal M}^*_4}]   +
 (M_Z^2 - t )^2 [ \vert  {{\cal M}_3 \vert }^2 +  \vert {{\cal M}_4}
\vert ^2]  \nonumber  \\
[0.2cm] & &
 +   4 (M_Z^2 - u) Re[ {{\cal M}_1} {{\cal M}^*_5} +
{{\cal M}_2} {{\cal M}^*_6}]
+  (M_Z^2 - u)^2 [  \vert {{\cal M}_5 \vert }^2 +
\vert {{\cal M}_6 \vert }^2]
\nonumber \\ [0.2cm] & &
 - 2 (M_{A}^2 M_Z^2 + M_Z^2 s - t u) Re[ {{\cal M}_3}  {{\cal M}^*_5} +
 {{\cal M}_4}  {{\cal M}^*_6} ]  \}
\end{eqnarray}
The differential cross--section reads:
\begin{eqnarray}
\frac{d \sigma}{d \Omega}(e^+ e^- \to Z A^0)= 
\frac{1}{256 \pi^2 s^2} 
\sqrt{(s-(M_{A} + M_Z)^2)(s-(M_{A} -M_Z)^2)}
\sum_{Z Pol}
\vert {\cal M}^1 \vert^2 .
\end{eqnarray}
In the on--shell scheme defined in \cite{dh},
it is found that  
there is no counter--term for the vertices Z-Z-$A^0$, 
Z-$\gamma$-$A^0$ and  $\gamma$-$\gamma$-$A^0$. 
Consequently the one--loop vertices
Z-Z-$A^0$, Z-$\gamma$-$A^0$   and $\gamma$-$\gamma$-$A^0$ 
have to be separately UV finite.
It can be seen from the expression of the invariant 
${\cal A}_i$ (eqs. \ref{ai}) that the operators 
${\cal A}_{1,2}$ are dimension 3 while 
operators ${\cal A}_{3,4,5,6}$ are dimension 5. 
One concludes that ${\cal M}_{3,4,5,6}$ have to be UV finite 
while ${\cal M}_{1,2}$ can potentially be UV divergent. 
We have checked numerically and 
analytically that the full set of invariants ${\cal M}_{i}$
are UV finite as required, and this feature will 
provide us with a good check of our calculation.

\renewcommand{\theequation}{3.\arabic{equation}}
\setcounter{equation}{0}
\section*{3. Numerical results and discussion}
In this section we focus on
the numerical analysis.
We take the fine structure constant in the Thomson limit: 
$\alpha=1/137.03598$.

\begin{figure}[t!]
\smallskip\smallskip 
\centerline{{
\epsfxsize2.8 in 
\epsffile{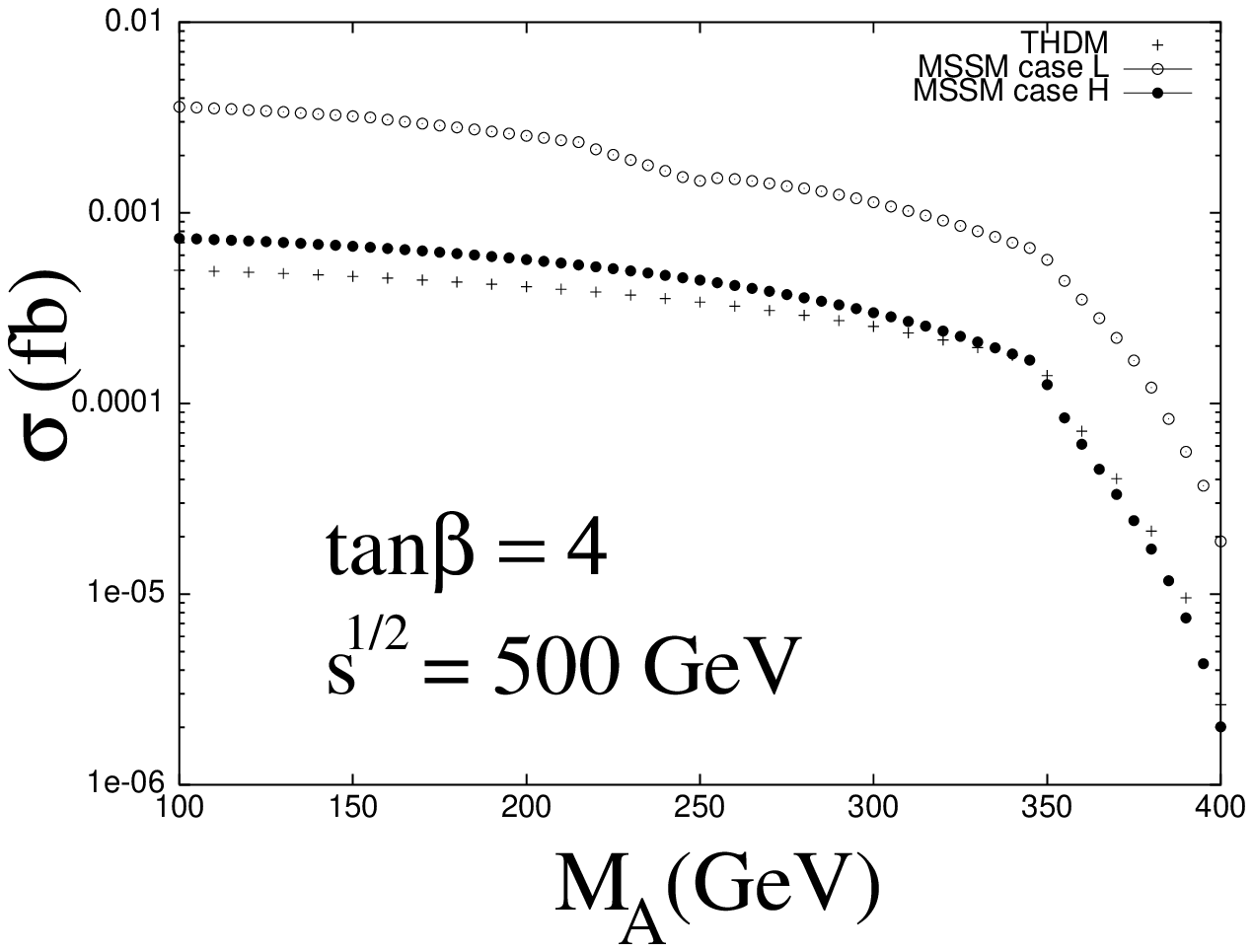}}
\hskip0.4cm
{\epsfxsize2.8 in \epsffile{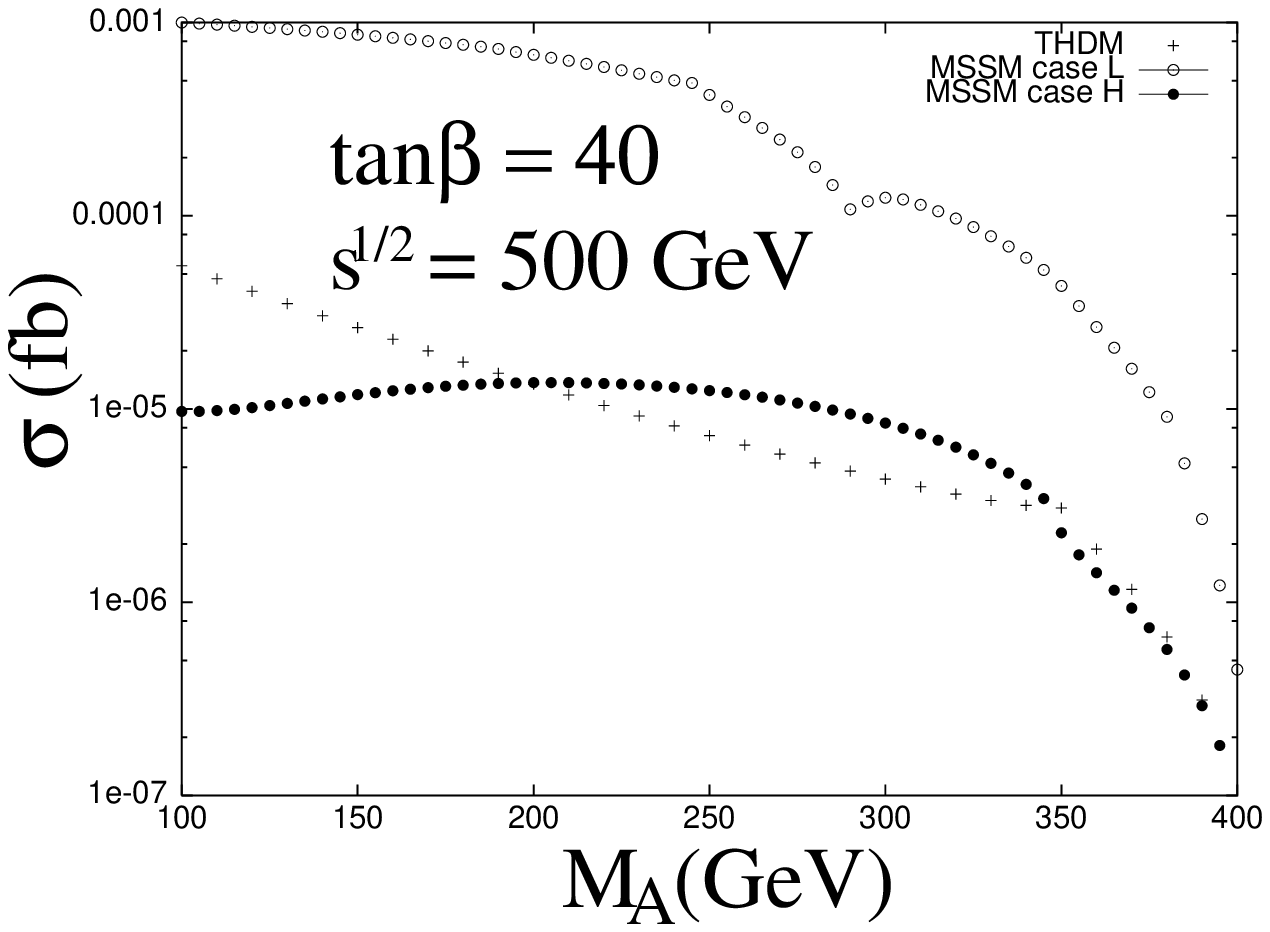}}}
\smallskip\smallskip
\caption{ Total cross--section at $500$ GeV
 centre of mass energy for 
$\tan\beta=4$ 
(left) and $\tan\beta=40$ (right)}
\label{cros1}
\end{figure}

The MSSM Higgs sector is parametrized by the mass of 
CP--odd $M_A$ and $\tan\beta$ while the top quark mass and 
the associated squark masses enter through radiative 
corrections \cite{okada}. In our study we 
will include the leading corrections only, where the light Higgs mass
is given by:
\begin{eqnarray} 
& & m^2_{h^0} =  \frac{1}{2}\Big[ m_{AZ}^2
- \sqrt{ m_{AZ}^4-4m_A^2 m_Z^2\, {\cos}^2\left( 2\beta \right)
-4\epsilon \left( m_A^2 \,s_\beta^2 +m_Z^2 \, c_\beta^2 \right) } \Big] .
\end{eqnarray}
with 
\begin{eqnarray}
& & m_{AZ}^2 = m_A^2+m_Z^2+\epsilon \qquad , 
\qquad  \epsilon = \frac{3 G_F}{\sqrt{2} \pi^2 } \frac{m_t^4}{ s_\beta^2}
\mbox{log} \left[\frac{m_{\tilde{t}}^2}{m_t^2} \right] 
\end{eqnarray}
The parameter $m_{  \tilde{t}}^2=m_{\tilde{t}_1}m_{\tilde{t}_2}$ denotes
the average mass squared of the stop particles.
A recent search for $h^0$ and $A^0$ excludes the region 
$0.5< \tan\beta <2.4$ \cite{www}, assuming maximal mixing in the stop sector.
In light of this we limit ourselves to the case where $\tan\beta\geq 3$.
Following the approach of \cite{gun} we assume a detection 
threshold of 0.1 fb for $e^+e^-\to A^0Z$. This would give 50 events before
experimental cuts for the expected luminosities of 500 fb$^{-1}$. 
In the THDM this criterion would require $\tan\beta\leq 0.3$, even for a 
light pseudoscalar \cite{AAC}. 

The chargino/neutralino sector 
can be parametrized by the usual $M_1$, $M_2$ and $\mu$.
We limit ourselves to the case of $\mu >0$ since
$\mu < 0$ seems to be disfavoured by $b\to s\gamma $ 
constraint\footnote{the chargino-stop contribution 
to $b\to s \gamma$ add constructively with the SM 
one for $\mu <0$ \cite{bsg}.} and from recent measurements of
the anomalous magnetic moment of the muon, $(g-2)_{\mu}$ \cite{martin}. 
We also use the SUGRA
constraint $M_1\approx M_2/2$ and take the left selectron, right 
selectron and sneutrino to be degenerate with a common mass $M_{\tilde{l}}$

We present our results for two specific cases: (i)  Light SUSY
case (case L), where all sleptons, charginos and neutralinos are 
relatively light with mass $\leq 200$ GeV. (ii) Heavy SUSY case
(case H) where all sleptons, charginos and neutralinos have mass 
$\geq 450 $ GeV.
\\
\\
$\bullet$ Case L: $M_2=220$, $\mu= 180$ and $M_{\tilde{l}}=200$ GeV\\
$\bullet$ Case H: $M_2=500$, $\mu= 600$ and $M_{\tilde{l}}=450$ GeV
\\

We start by recalling that the THDM contribution
to $e^+e^-\to A^0Z$ \cite{AAC} in the small $\tan \beta$
regime is enhanced by the top quark contribution, leading to 
significant cross--sections (about a few fb). In the large 
$\tan \beta$ regime the cross--section is suppressed and does 
not attain observable rates. 
In the MSSM we limit ourself to the case where $\tan\beta\geq 3$,
and consequently the THDM contribution is suppressed to the order 
of $\approx 0.001$ fb at $\sqrt s=500$ GeV.
Our aim is to see if the SUSY contribution 
can enhance the cross--section.

In Fig.2 we show the dependence of the total cross--section 
on the CP--odd Higgs mass $M_A$ for $\sqrt{s}=500$ GeV 
in the case where $\tan\beta=4$ (Fig.3.a) and 40 (Fig.3.b). 
In both cases we show the MSSM contribution for case L and case H,
as well as the THDM contribution. It can be seen that the light SUSY 
particle scenario (L) may enhance the total cross--section,
which can reach $\approx 0.003$ fb for $M_A\leq 300$ GeV 
and $\tan\beta=4$, with smaller cross--sections for $\tan\beta=40$. 
Note also that for $M_A\leq 300$ GeV the box 
contribution is an order of magnitude smaller than the vertex 
corrections in the MSSM case L, for both $\tan\beta=4$  and 40. 
The kink observed in this figure
close to $M_A\approx 2m_t$ is a threshold effect
due to the opening of the decay $A^0 \to t\bar{t}$. 
One can also have other threshold effects which arise from 
charginos and neutralinos. Note that in the 
case L and for $\tan\beta=4$ (resp 40) the mass of the light 
chargino is about 132.5 GeV (resp 146 GeV), and
one can see a smooth threshold effect due to the opening of the channel 
$A^0\to \widetilde{\chi}_1^+ \widetilde{\chi}_1^-$ 
for $M_A\approx 264$ GeV (resp $M_A\approx 290$ GeV).
Due to the decoupling property of SUSY theory,
the MSSM contributions in case H are very small and 
the total cross--section is close to the THDM one.

\begin{figure}[t!]
\smallskip\smallskip 
\centerline{{
\epsfxsize2.8 in 
\epsffile{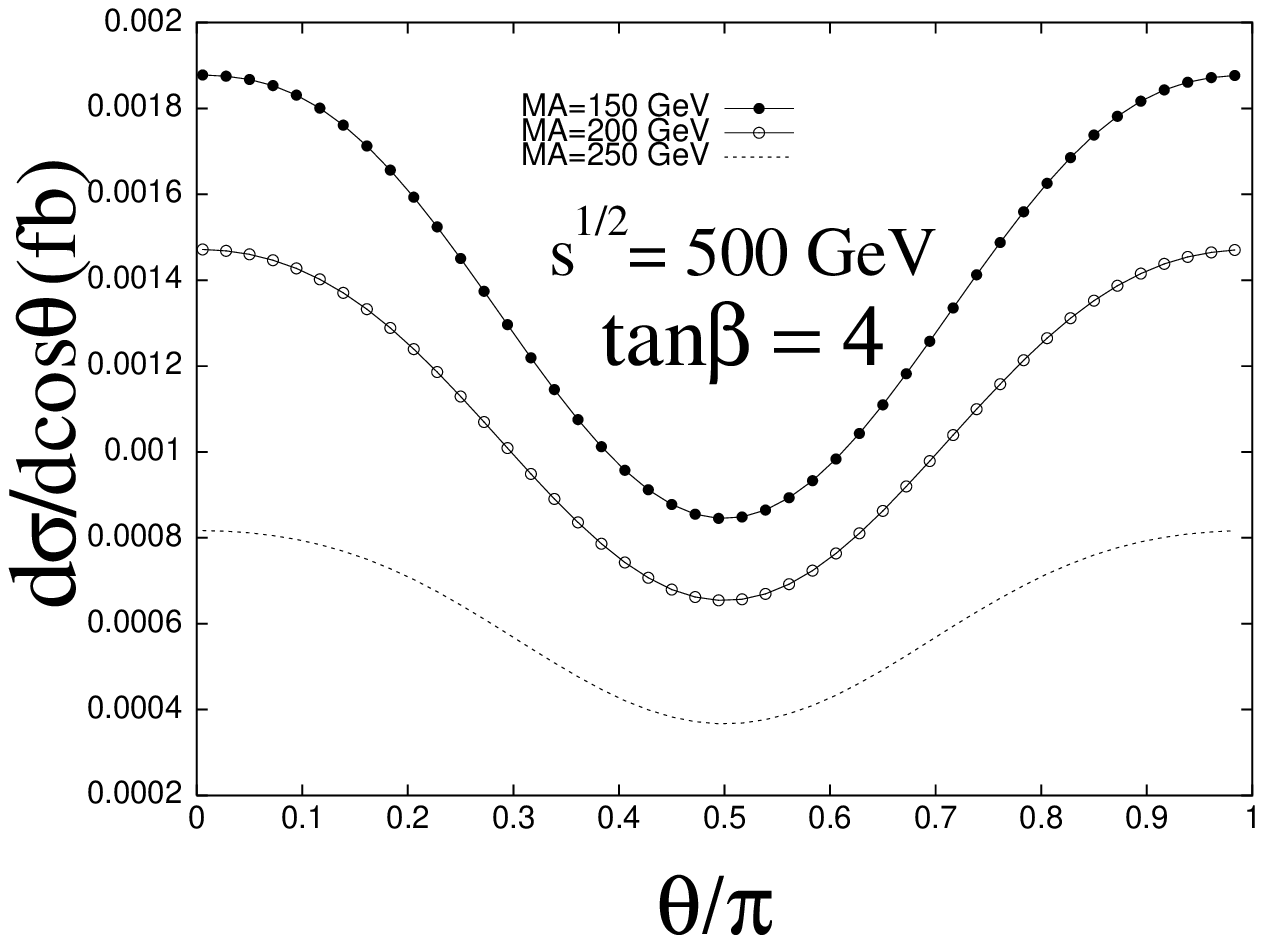}}
\hskip0.4cm
            {\epsfxsize2.8 in \epsffile{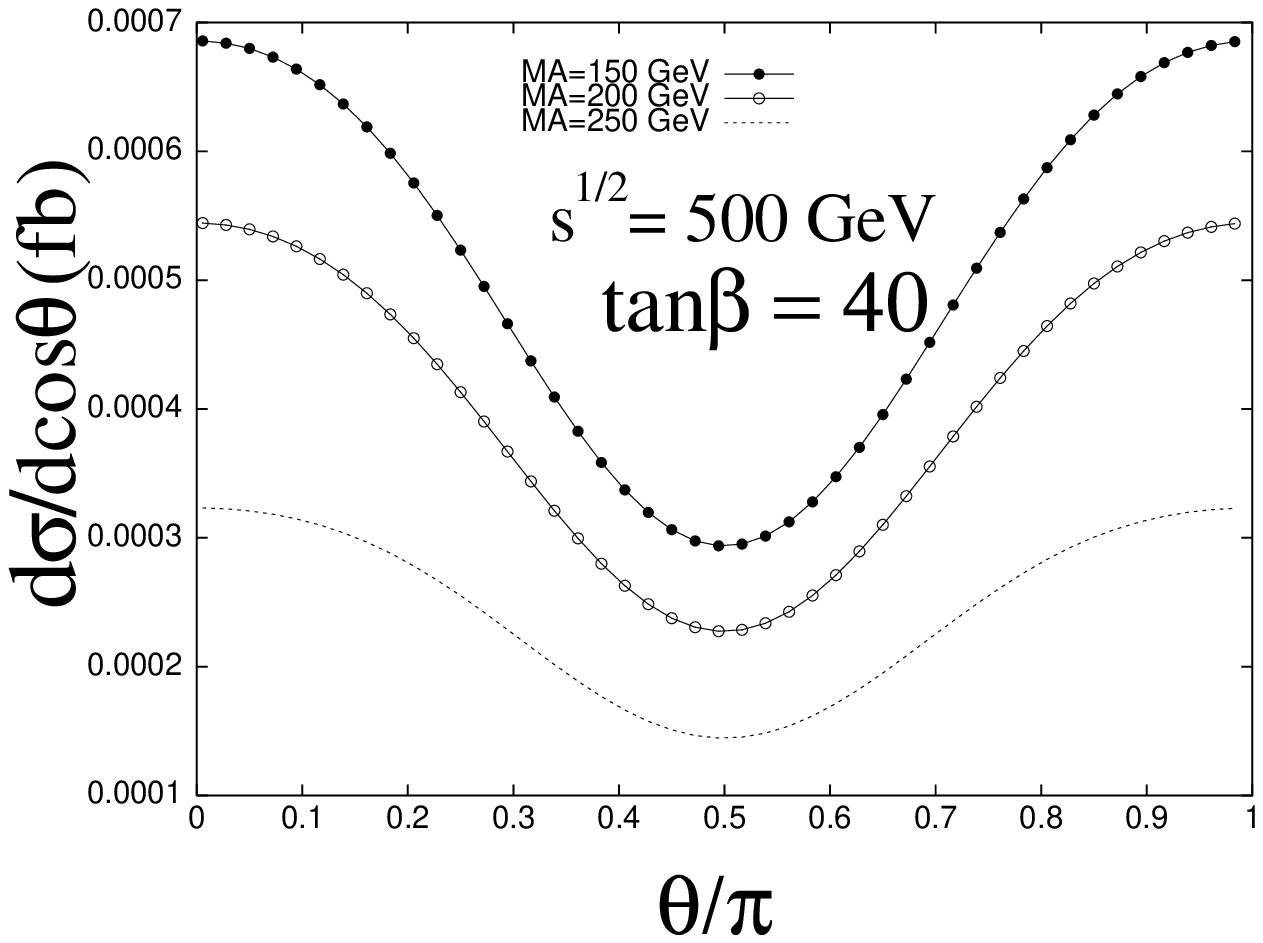}}}
\smallskip\smallskip
\caption{Angular distribution $\frac{d\sigma}{d\cos\theta}$ (fb) 
for $e^+e^- \to ZA^0$, MSSM case L with $\tan\beta=4$ (left) and 
$\tan\beta=40$ (right) }
\label{cros2}
\end{figure}


We illustrate in Fig.3 the angular distribution 
in the MSSM with light SUSY particles for $\sqrt{s}=500$ GeV,
$\tan\beta=4$ (left curve)
and $\tan\beta=40$ (right curve). 
The shape of the graph is in agreement with that expected for
a CP--odd scalar being produced by an effective $ZZA^0$ coupling
\cite{distrib}.

We stress here that in both cases, the shape of
the angular distribution originates from the vertex contribution,
since the box contribution is rather small at $\sqrt s=500$ GeV.
In the case of heavy SUSY particles (MSSM case H)
the total cross--section is suppressed which would render difficult
any analysis of the angular distribution. The numerical value of 
$\frac{d\sigma}{d\cos\theta}$ lies in the range
0.0019-0.0009 fb (resp 0.0008-0.0004 fb) with $0<\theta <\pi$
for $\tan\beta=4$ and $M_A=150$ GeV (resp $M_A=250$ GeV).
We note in passing that the corresponding values for 
the THDM with $\tan\beta\geq 3 $ would also be very small.

It is interesting to study the behaviour of the total 
cross--section versus $\sqrt{s}$. Since
a variety of SUSY particles are exchanged in the vertices and boxes,  
we expect that more threshold effects will appear. 
In Fig.4, we present the total cross--section against $\sqrt{s}$ 
in the MSSM case L (left curve) and MSSM case H (right curve)
for several values of $M_A$. 
It can be seen from the plot that there is a peak in both cases.
In the MSSM case L, the peak appears at $\sqrt{s}\approx 400$ GeV
since in this case the masses of SUSY particles are approximately 200 GeV.
In MSSM case H, the peak shows up for $\sqrt{s}\approx 1$ TeV. 
As one can see, the lighter $M_A$ is, the more spectacular are
the peaks. In the MSSM case L, for $\sqrt{s}<1.2$ TeV  
the dominant contribution to the total cross--section is from 
vertex diagrams, while for energy $\sqrt{s}>1.2$ TeV
it is the box diagrams which dominate.
Similarly in the MSSM case H, the vertex contibution
dominates for $\sqrt{s}<2.2$ TeV while the box diagram dominate if
$\sqrt{s}> 2.2$ TeV.  

\begin{figure}[t!]
\smallskip\smallskip 
\centerline{{
\epsfxsize2.8 in 
\epsffile{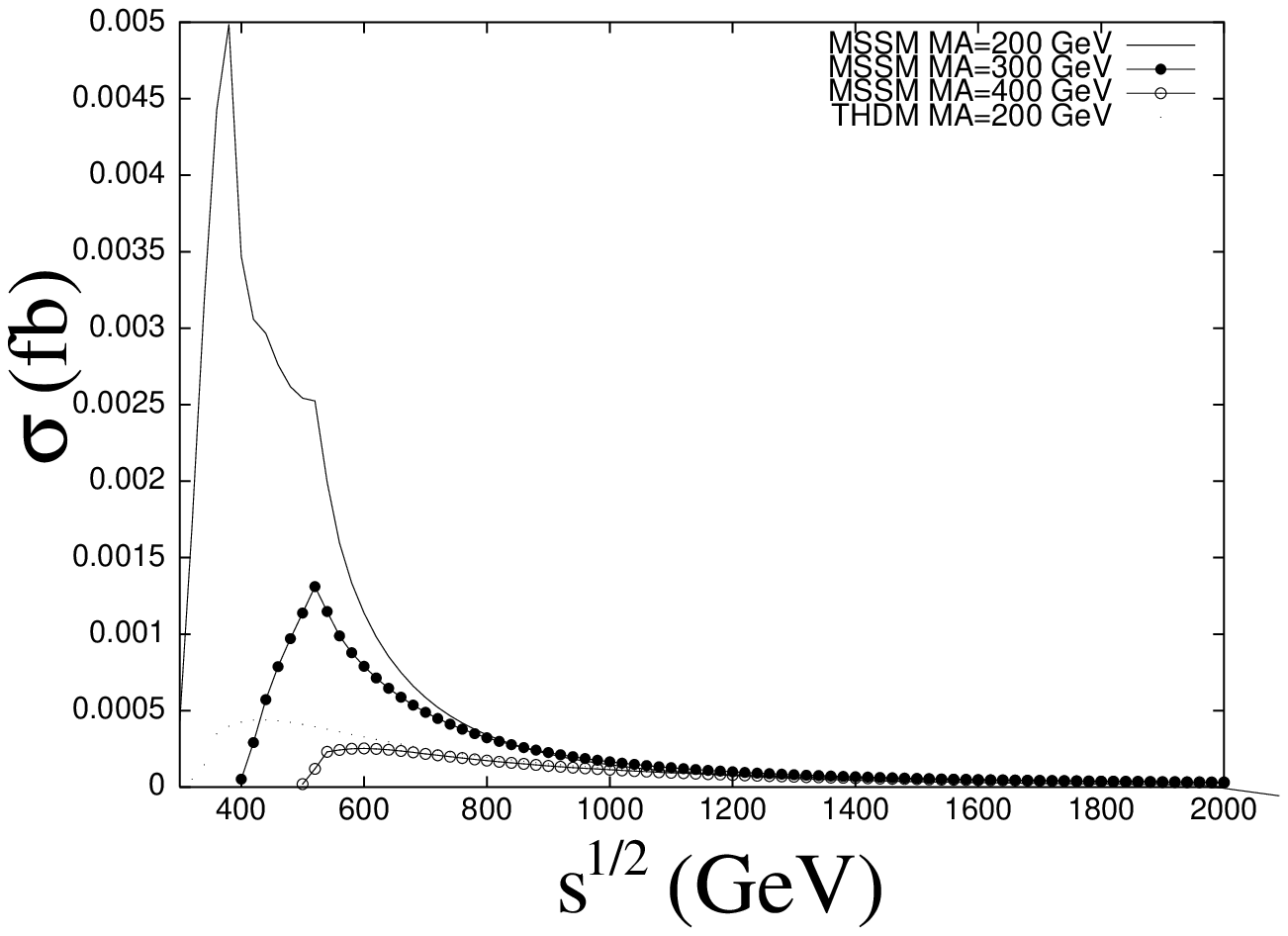}}
\hskip0.4cm
            {\epsfxsize2.8 in \epsffile{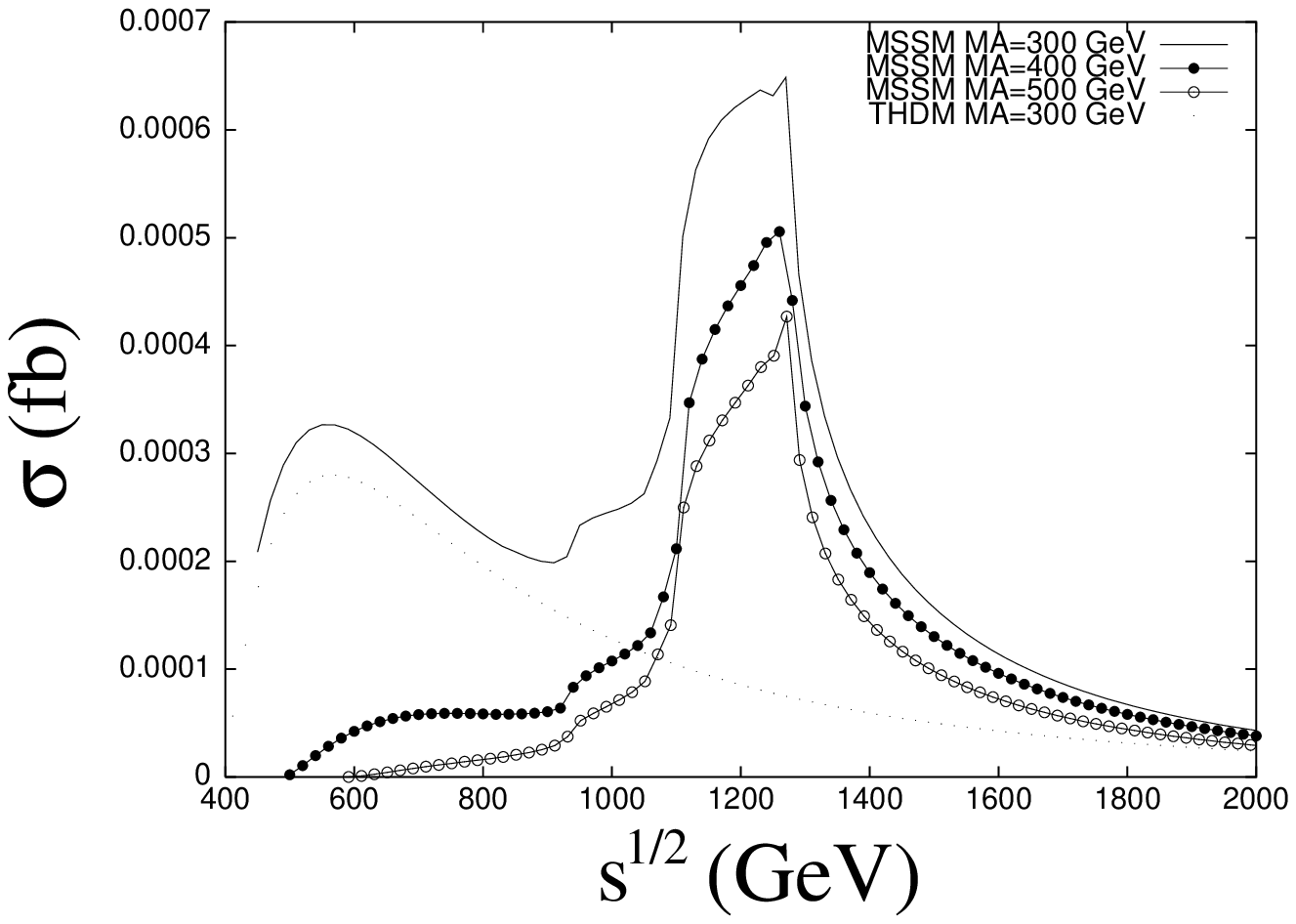}}}
\smallskip\smallskip
\caption{Total cross section $\sigma$ (fb) 
for $e^+e^- \to Z A$, MSSM case L with $\tan\beta=4$ (left) and 
MSSM case H  ($\tan\beta=4$ right) }
\label{cros2}
\end{figure}

\renewcommand{\theequation}{4.\arabic{equation}}
\setcounter{equation}{0}
\section*{4. Summary}
We have computed the cross--section for the production
mechanism $e^+e^-\to A^0Z$ at high energy $e^+e^-$ colliders
in the framework of the MSSM. Such a process proceeds via higher order
diagrams and is strongly model dependent. 

The calculation was performed within the dimensional regularisation scheme
and we presented results for both the MSSM and THDM. In the former case 
light SUSY particles may give important contributions
to the cross--section, resulting in maximum values 
of order 0.003 fb. 
Therefore the SUSY enhancement is not sufficient to produce
an observable signal at the planned luminosities of $500 fb^{-1}$. In
the MSSM with explicit CP violating phases, the
pseudoscalar contains a CP even component and may be produced
with an observable cross-section \cite{PRD64}. Therefore signals
in the Higgsstrahlung channel for all $h^0$, $H^0$ and $A^0$ could not be
explained in the MSSM unless SUSY sources of CP violation are present.
Observation of this process might enable one to 
distinguish between SUSY and non--SUSY Higgs sectors. 
We showed that threshold effects may enhance the cross--section. 

\vspace{1.cm}

{\bf Acknowledgments:} 
We thank F. Renard for useful discussions and for reading
the manuscript.
A. Arhrib  is supported  by NSC-89-2112-M-002-0129, Taiwan R.O.C.

\renewcommand{\theequation}{B.\arabic{equation}}
\setcounter{equation}{0}

\end{document}